\documentstyle[12pt]{article}

\def\noi{\noindent}

\def\nqq{\hspace{-2em}}

\def\barr{\left(\begin{array}}
\def\earr{\end{array}\right)}
\def\beq#1{\begin{equation}\label{#1}}
\def\eeq{\end{equation}}
\def\nn{\nonumber}
\def\mm{\\ \nqq}
\newcommand{\bear}[1]{\begin{eqnarray}\label{#1}}
\newcommand{\ear}{\end{eqnarray}}

\catcode`\@=11 \@addtoreset{equation}{section}\catcode`\@=12

\newcommand{\R}{\mbox{\bf R}}
\newcommand{\N}{\mbox{\bf N}}

\newcommand{\sign}{\mathop{\rm sign}\nolimits}

\newcommand{\eps}{\varepsilon}
\newcommand{\tri}{\triangle}
\newcommand{\p}{\partial}

\newcommand{\fnm}{\footnotemark}
\newcommand{\fnt}{\footnotetext}

\begin{document}

\begin{center}
\large\bf

COMPOSITE FLUXBRANES WITH GENERAL INTERSECTIONS
\\[15pt]
\normalsize\bf V.D. Ivashchuk\fnm[1]\fnt[1]{ivas@rgs.phys.msu.su},\\

\it Center for Gravitation and Fundamental Metrology,
VNIIMS, 3/1 M. Ulyanovoy Str., Moscow 119313, Russia  \\

Institute of Gravitation and Cosmology,
Peoples' Friendship University of Russia, \\
6 Miklukho-Maklaya Str.,  Moscow 117198, Russia

\end{center}

\vspace{15pt}

\small\noi

\begin{abstract}

Generalized composite fluxbrane  solutions  for
a wide class of intersection rules are obtained.
The solutions are defined on a manifold which contains a product
of $n$ Ricci-flat spaces $M_1 \times \ldots \times M_n$ with
1-dimensional $M_1$.
They are defined up to  a set of functions $H_s$ obeying
non-linear differential equations
equivalent to Toda-type equations with
certain boundary conditions imposed. A conjecture on polynomial
structure of governing functions $H_s$ for intersections related to
semisimple Lie algebras is suggested.
This conjecture is valid for Lie algebras: $A_m$, $C_{m+1}$, $m \geq 1$.
For simple Lie algebras
the powers of polynomials coincide with the components  of the dual Weyl
vector in the basis of simple roots.
Explicit formulas for $A_1 \oplus \ldots \oplus A_1$ (orthogonal),
"block-ortogonal" and $A_2$ solutions are obtained.  Certain examples of
solutions in $D = 11$ and $D =10$ ($II A$) supergravities
(e.g.  with $A_2$ intersection rules) and
Kaluza-Klein  dyonic $A_2$ flux tube, are considered.

\end{abstract}

\hspace{1cm}*PACS numbers:
0420J, 0450, 0470

\pagebreak

\normalsize

%%%%%%%%%%%%%%%%%%%%%%%%%%%%%%%%%%%%%%%%%%%%%%%%%%%%%%%%%%%%%%%%
\section{Introduction}
%%%%%%%%%%%%%%%%%%%%%%%%%%%%%%%%%%%%%%%%%%%%%%%%%%%%%%%%%%%%%%%%

Recently a lot of papers were devoted to multidimensional
generalizations of the well-known Melvin solution \cite{Melv}
(for exact solutions and their applications see
\cite{GM}-\cite{EmpGut} and references therein).
We remind that the
original Melvin solution describes the gravitational field
of a magnetic flux tube.
In the works of Gibbons and Wiltshire \cite{GW}
and Gibbons and Maeda \cite{GM}
the Melvin solution was generalized to  arbitrary dimensions
(in \cite{GM} with the inclusion of dilaton), and hence the simplest
fluxbranes appeared.
\fnm[2]\fnt[2]{The reference \cite{GW} is added in the new version}.
(A term fluxbrane was suggested in \cite{DGGH2}.
The Melvin solution is currently denoted as $F1$.)
In papers  \cite{DGKT,DGGidH,DGGH1,DGGH2},
devoted to Kaluza-Klein Melvin solution (e.g. non-perturbative
instability and pair production of magnetically charged black holes)
it was shown that $F1$-fluxbrane supported
by potential $1$-form has a nice interpretation
as a modding of flat space in one dimension higher.
This "modding" technique is widely used in construction of
new solutions in supergravitational models and also
for various physical applications in string and $M$-theory:
construction of exact string backgrounds \cite{RTs1,Tseyt},
duality between $0A$ and $IIA$  string theories \cite{CosG},
dielectric effect  \cite{CosHC,Emp,BSaf}, construction of
supersymmetric configurations \cite{FFS,RTs3} etc

The "modding"  technique may be also used for $Fp$-fluxbranes
supported by forms of higher ranks (constructed by
 use of $1$-forms). Another approach suggested in \cite{GR,CGS},
which works for $Fp$-fluxbranes, is based on generating techniques
for certain duality groups. An important result here is
a construction by Chen, Gal'tsov and Sharakin \cite{CGS}
of intersecting $F6$ and $F3$ fluxbranes corresponding to $M$-branes
in $D=11$ supergravity.

The third and the most direct method is based on solving
of Einstein equations \cite{Saf1,GutSt,CGSaf}. This article is
devoted to further developing of this approach
using the $p$-brane solutions from \cite{IK}
(for a review of $p$-brane solutions see \cite{St,IMtop}).
We remind, that in \cite{IK} a family of $p$-brane  "cosmological-type"
solutions with nearly arbitrary  (up to some restrictions) intersection
rules  were obtained (see Sect. 2). These solutions are defined up to
solutions to Toda-type equations and contain as a special case a subclass
of solutions with cylindrical symmetry.
Here we single out a subclass of generalized fluxbrane configurations
related to Toda-type equations with  certain asymptotical conditions
imposed (Sect. 3).  These fluxbrane solutions are governed by functions
$H_s(z) > 0$ defined on the interval $(0, z_0)$
and obeying a set of second order non-linear differential equations
\beq{0.1}
 \frac{d}{dz} \left( \frac{ z}{H_s} \frac{d}{dz} H_s \right) =
 \hat B_s \prod_{s' \in S}  H_{s'}^{- A_{s s'}},
\eeq
with  the following boundary conditions imposed:
\begin{eqnarray}
\label{0.2}
{\bf (i)} \quad H_{s}(+ 0) = 1,
\end{eqnarray}
$s \in S$, ($S$ is non-empty set). In (\ref{0.1}) all
$\hat B_s \neq 0$ are constants
and  $(A_{s s'})$ is a "quasi-Cartan" matrix
($A_{ss} = 2$)
coinciding with the Cartan one when intersections are related to Lie
algebras. In most interesting examples  $z_0 = + \infty$
and all $\hat B_s > 0$.

We remind that a slightly different equations occur
for black brane solutions from \cite{IMp1,IMp2,IMp3}
\beq{0.3}
 \frac{d}{dz} \left( \frac{(1 - 2\mu z)}{H_s} \frac{d}{dz} H_s \right) =
 \bar B_s \prod_{s' \in S}  H_{s'}^{- A_{s s'}},
\eeq
where $\mu > 0$ is extremality parameter,
$z \in (0, (2\mu)^{-1})$ and all $\bar B_s \neq 0$,
(in most interesting cases $\bar B_s < 0$).
For black brane solutions the finite limits (on a horizon)
$H_{s}((2\mu)^{-1} -0) = H_{s0} > 0$ exist.
We note that fluxbrane "master equations" (\ref{0.1})  may be
obtained from the black brane ones (\ref{0.3})
in the limit $\mu \to \infty$. For different aspects
of p-brane/fluxbrane correspondence see \cite{Saf2,EmpGut}.

In Sect. 4 we suggest a hypothesis: the solutions
to eqs. (\ref{0.1}), (\ref{0.2})
are polynomials when intersection rules correspond to
semisimple Lie algebras.  This hypothesis ({\bf Conjecture })
may be proved for Lie algebras $A_m$, $C_{m+1}$, $m = 1,2, \ldots$,
along a line as it was done in \cite{IMp2,IMp3} for
black brane solutions. It is also confirmed by special "block-orthogonal"
fluxbrane solutions and explicit formulas for
the solutions corresponding to  Lie algebras
$A_1 \oplus \ldots \oplus A_1$ and $A_2$.
In Sect. 5 certain examples of
solutions in $D = 11$ supergravity, i.e. $F6$ and $F3$ fluxbranes,
$F6 \cap F3$  solutions with $A_1 \oplus A_1$ and $A_2$ intersection rules,
$F7,(NS)F6,F5$ and (dual) $F1, (NS) F2, F3$ fluxbranes
in $D = 10$ $IIA$ supergravity and Kaluza-Klein
dyonic $A_2$-flux tube, are considered.

%%%%%%%%%%%%%%%%%%%%%%%%%%%%%%%%%%%%%%%%%%%%%%%%%%%%%%%%%%%%%%%%
\section{$p$-brane cosmological-type solutions \\
with general intersection rules }
%%%%%%%%%%%%%%%%%%%%%%%%%%%%%%%%%%%%%%%%%%%%%%%%%%%%%%%%%%%%%%%%

\subsection{The model}

We consider a  model governed by the action \cite{IMC}
\beq{1.1}
S=\int d^Dx \sqrt{|g|}\biggl\{R[g]-h_{\alpha\beta}g^{MN}\p_M\varphi^\alpha
\p_N\varphi^\beta-\sum_{a\in\tri}\frac{\theta_a}{n_a!}
\exp[2\lambda_a(\varphi)](F^a)^2\biggr\}
\eeq
where $g=g_{MN}(x)dx^M\otimes dx^N$ is a metric,
$\varphi=(\varphi^\alpha)\in\R^l$ is a vector of scalar fields,
$(h_{\alpha\beta})$ is a  constant symmetric
non-degenerate $l\times l$ matrix $(l\in \N)$,
$\theta_a=\pm1$,
\beq{1.2a}
F^a =    dA^a
=  \frac{1}{n_a!} F^a_{M_1 \ldots M_{n_a}}
dz^{M_1} \wedge \ldots \wedge dz^{M_{n_a}}
\eeq
is a $n_a$-form ($n_a\ge1$), $\lambda_a$ is a
1-form on $\R^l$: $\lambda_a(\varphi)=\lambda_{\alpha a}\varphi^\alpha$,
$a\in\tri$, $\alpha=1,\dots,l$.
In (\ref{1.1})
we denote $|g| =   |\det (g_{MN})|$,
\beq{1.3a}
(F^a)^2_g  =
F^a_{M_1 \ldots M_{n_a}} F^a_{N_1 \ldots N_{n_a}}
g^{M_1 N_1} \ldots g^{M_{n_a} N_{n_a}},
\eeq
$a \in \tri$. Here $\tri$ is some finite set.

\subsection{"Cosmological-type" solutions}

Let us consider a family of
solutions to field equations corresponding to the action
(\ref{1.1}) and depending upon one variable $u$
\cite{IK}. These solutions are defined on the manifold
\beq{1.2}
M =    (u_{-}, u_{+})  \times
M_1  \times M_2 \times  \ldots \times M_{n},
\eeq
where $(u_{-}, u_{+})$  is  an interval belonging to $\R$.
The solutions read \cite{IK}
\bear{1.3}
g= \biggl(\prod_{s \in S} [f_s(u)]^{2 d(I_s) h_s/(D-2)} \biggr)
\biggr\{ \exp(2c^0 u + 2 \bar c^0) w du \otimes du  + \\ \nn
\sum_{i = 1}^{n} \Bigl(\prod_{s\in S}
[f_s(u)]^{- 2 h_s  \delta_{i I_s} } \Bigr)
\exp(2c^i u+ 2 \bar c^i) g^i \biggr\}, \\ \label{1.4}
\exp(\varphi^\alpha) =
\left( \prod_{s\in S} f_s^{h_s \chi_s \lambda_{a_s}^\alpha} \right)
\exp(c^\alpha u + \bar c^\alpha), \\ \label{1.5}
F^a= \sum_{s \in S} \delta^a_{a_s} {\cal F}^{s},
\ear
$\alpha=1,\dots,l$.
In  (\ref{1.3})  $w = \pm 1$,
$g^i=g_{m_i n_i}^i(y_i) dy_i^{m_i}\otimes dy_i^{n_i}$
is a Ricci-flat  metric on $M_{i}$, $i=  1,\ldots,n$,
\beq{1.11}
\delta_{iI}=  \sum_{j\in I} \delta_{ij}
\eeq
is the indicator of $i$ belonging
to $I$: $\delta_{iI}=  1$ for $i\in I$ and $\delta_{iI}=  0$ otherwise.

The  $p$-brane  set  $S$ is by definition
\beq{1.6}
S=  S_e \sqcup S_m, \quad
S_v=  \sqcup_{a\in\tri}\{a\}\times\{v\}\times\Omega_{a,v},
\eeq
$v=  e,m$ and $\Omega_{a,e}, \Omega_{a,m} \subset \Omega$,
where $\Omega =   \Omega(n)$  is the set of all non-empty
subsets of $\{ 1, \ldots,n \}$.
Here and in what follows $\sqcup$ means the union
of non-intersecting sets. Any $p$-brane index $s \in S$ has the form
\beq{1.7}
s =   (a_s,v_s, I_s),
\eeq
where
$a_s \in \tri$ is colour index, $v_s =  e,m$ is electro-magnetic
index and the set $I_s \in \Omega_{a_s,v_s}$ describes
the location of $p$-brane worldvolume.

The sets $S_e$ and $S_m$ define electric and magnetic
$p$-branes correspondingly. In (\ref{1.4})
\beq{1.8}
\chi_s  =   +1, -1
\eeq
for $s \in S_e, S_m$ respectively.
In (\ref{1.5})  forms
\beq{1.9}
{\cal F}^s= Q_s
\left( \prod_{s' \in S}  f_{s'}^{- A_{s s'}} \right) du \wedge\tau(I_s),
\eeq
$s\in S_e$, correspond to electric $p$-branes and
forms
\beq{1.10}
{\cal F}^s= Q_s \tau(\bar I_s),
\eeq
$s \in S_m$,
correspond to magnetic $p$-branes; $Q_s \neq 0$, $s \in S$.
In (\ref{1.10})  and in what follows
\beq{1.13a}
\bar I \equiv I_0 \setminus I, \qquad I_0 = \{1,\ldots,n \}.
\eeq

All manifolds $M_{i}$ are assumed to be oriented and
connected and  the volume $d_i$-forms
\beq{1.12}
\tau_i  \equiv \sqrt{|g^i(y_i)|}
\ dy_i^{1} \wedge \ldots \wedge dy_i^{d_i},
\eeq
  and parameters
 \beq{1.12a}
 \varepsilon(i)  \equiv {\rm sign}( \det (g^i_{m_i n_i})) = \pm 1
 \eeq
%\fnm[3]\fnt[3]{This  relation was missed in the old version.}
are well--defined for all $i=  1,\ldots,n$.
Here $d_{i} =   {\rm dim} M_{i}$, $i =   1, \ldots, n$,
$D =   1 + \sum_{i =   1}^{n} d_{i}$. For any
 $I =   \{ i_1, \ldots, i_k \} \in \Omega$, $i_1 < \ldots < i_k$,
we denote
\bear{1.13}
\tau(I) \equiv \tau_{i_1}  \wedge \ldots \wedge \tau_{i_k},
\\ \label{1.14}
M(I) \equiv M_{i_1}  \times  \ldots \times M_{i_k}, \\
\label{1.15}
d(I) \equiv {\rm dim } M(I) =  \sum_{i \in I} d_i,
\ear
 \begin{equation}
 \varepsilon(I) \equiv \varepsilon(i_1) \ldots \varepsilon(i_k).
 \end{equation}
%\fnm[4]\fnt[4]{The last relation was missed in the old version.}
$M(I_s)$ is isomorphic to $p$-brane worldvolume
(see (\ref{1.7})).

The parameters  $h_s$ appearing in the solution
satisfy the relations
\beq{1.16}
h_s = K_s^{-1}, \qquad  K_s = B_{s s},
\eeq
where
\beq{1.17}
B_{ss'} \equiv
d(I_s\cap I_{s'})+\frac{d(I_s)d(I_{s'})}{2-D}+
\chi_s\chi_{s'}\lambda_{\alpha a_s}\lambda_{\beta a_{s'}}
h^{\alpha\beta},
\eeq
$s, s' \in S$, with $(h^{\alpha\beta})=(h_{\alpha\beta})^{-1}$.
Here we assume that
\beq{1.17a}
({\bf i}) \qquad B_{ss} \neq 0,
\eeq
for all $s \in S$, and
\beq{1.18b}
({\bf ii}) \qquad {\rm det}(B_{s s'}) \neq 0,
\eeq
i.e. the matrix $(B_{ss'})$ is a non-degenerate one. In (\ref{1.9})
another non-degenerate matrix (a so-called "quasi-Cartan" matrix)
appears
\beq{1.18}
(A_{ss'}) = \left( 2 B_{s s'}/B_{s' s'} \right).
\eeq
Here  some ordering in the set $S$ is assumed.

In (\ref{1.3}), (\ref{1.4})
\beq{1.19}
f_s = \exp( - q^s),
\eeq
where $(q^s) = (q^s(u))$ is a solution to Toda-type equations
\beq{1.20}
\ddot{q^s} = -  B_s \exp( \sum_{s' \in S} A_{s s'} q^{s'} ),
\eeq
with
\beq{1.21}
 B_s = \eps_s K_s Q_s^2,
\eeq
$s \in S$. Here
\beq{1.22}
\eps_s=(-\eps[g])^{(1-\chi_s)/2}\eps(I_s) \theta_{a_s},
\eeq
$s\in S$, $\eps[g]\equiv\sign\det(g_{MN})$. More explicitly
(\ref{1.22}) reads: $\eps_s=\eps(I_s) \theta_{a_s}$ for
$v_s = e$ and $\eps_s=-\eps[g] \eps(I_s) \theta_{a_s}$, for
$v_s = m$.

Vectors $c=(c^A)= (c^i, c^\alpha)$ and
$\bar c=(\bar c^A)$ satisfy the linear constraints
\beq{1.27}
U^s(c)= \sum_{i \in I_s}d_ic^i-\chi_s\lambda_{a_s\alpha}c^\alpha=0,
\qquad
U^s(\bar c)=  \sum_{i\in I_s}d_i\bar c^i-
\chi_s\lambda_{a_s\alpha}\bar c^\alpha=0,
\eeq
$s\in S$,
and
\bear{1.30aa}
c^0 = \sum_{j=1}^n d_j c^j,
\qquad
\bar  c^0 = \sum_{j=1}^n d_j \bar c^j,
\\  \label{1.30a}
2E = 2E_{TL} + h_{\alpha\beta}c^\alpha c^\beta+ \sum_{i=1}^n d_i(c^i)^2
- \left(\sum_{i=1}^nd_ic^i\right)^2 = 0,
\ear
where
\beq{1.31}
E_{TL} = \frac{1}{4}  \sum_{s,s' \in S} h_s
A_{s s'} \dot{q^s} \dot{q^{s'}}
  + \sum_{s \in S} A_s  \exp( \sum_{s' \in S} A_{s s'} q^{s'} ),
\eeq
is an integration constant (energy) for the solutions from
(\ref{1.20}). Here $A_s =  \frac12  \eps_s Q_s^2= B_s h_s/2$,
$s \in S$. Eqs. (\ref{1.20}) correspond to the
Toda-type Lagrangian
\beq{1.31a}
L_{TL} = \frac{1}{4}  \sum_{s,s' \in S}
h_s  A_{s s'} \dot{q^s}\dot{q^{s'}}
-  \sum_{s \in S} A_s  \exp( \sum_{s' \in S} A_{s s'} q^{s'} ).
\eeq

{\bf Remark 1.}
{\em Here we identify notations  for $g^{i}$  and  $\hat{g}^{i}$, where
$\hat{g}^{i} = p_{i}^{*} g^{i}$ is the
pullback of the metric $g^{i}$  to the manifold  $M$ by the
canonical projection: $p_{i} : M \rightarrow  M_{i}$, $i = 1,
\ldots, n$. An analogous agreement will be also kept for volume forms etc.}

Due to (\ref{1.9}) and  (\ref{1.10}), the dimension of
$p$-brane worldvolume $d(I_s)$ is defined by
\beq{1.16a}
d(I_s)=  n_{a_s}-1, \quad d(I_s)=   D- n_{a_s} -1,
\eeq
for $s \in S_e, S_m$ respectively.
For a $p$-brane: $p =   p_s =   d(I_s)-1$.

\subsection{Restrictions on $p$-brane configurations.}

The solutions  presented above are valid if two
restrictions on the sets of $p$-branes are satisfied.
These  restrictions
guarantee  the block-diagonal form of the  energy-momentum tensor
and the existence of the sigma-model representation (without additional
constraints) \cite{IMC,IMtop}.

Let us denote
$w_1\equiv\{i|i\in\{1,\dots,n\},\quad d_i=1\}$, and
$n_1=|w_1|$ (i.e. $n_1$ is the number of 1-dimensional spaces among
$M_i$, $i=1,\dots,n$).

{\bf Restriction 1.} {\em For any $a\in\tri$ and $v= e,m$ there are no
$I,J \in\Omega_{a,v}$ such that
$ I= \{i\} \sqcup (I \cap J)$, and $J= (I \cap J) \sqcup \{ j \}$
for some $i,j \in w_1$, $i \neq j$.}

{\bf Restriction 2.}.
{\em For any $a\in\tri$ there are no
$I\in\Omega_{a,e}$ and $J\in\Omega_{a,m}$ such that
$\bar J=\{i\}\sqcup I$.}

Restriction 1  is satisfied for $n_1 \leq 1$ and also in
the non-composite case: $|\Omega_{a,e}|+ |\Omega_{a,m}| = 1$ for all
$a\in\tri$.  For $n_1\ge2$ it forbids the following
pairs of two electric or two magnetic $p$-branes,
corresponding to the same form $F^a, a \in \tri$:

\begin{center}
\begin{tabular}{cccc}
\cline{1-2}
\multicolumn{1}{|c|}{$i$} &
\multicolumn{1}{|c|}{\hspace*{1cm}} & & $\quad I$ \\
\cline{1-2}
 & & & \\
\cline{2-3}
 & \multicolumn{1}{|c|}{\hspace*{1cm}} &
\multicolumn{1}{|c|}{$j$} & $\quad J$ \\
\cline{2-3}
\end{tabular}
\end{center}

\begin{center}

{\bf \small Figure 1.
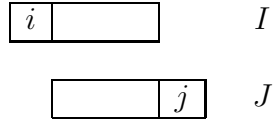
A forbidden by Restriction 1 pair of two electric or two
magnetic p-branes.
}

\end{center}

Here $d_i = d_j =1$, $i \neq j$, $i,j =1,\dots,n$. Restriction 1
may be also rewritten in terms of intersections
\beq{2.2.2a}
{\bf (R1)} \quad d(I \cap J) \leq d(I)  - 2,
\eeq
for any $I,J \in\Omega_{a,v}$, $a\in\tri$, $v= e,m$ (here $d(I) = d(J)$).

Restriction 2 is satisfied for $n_1=0$. For
$n_1\ge1$ it forbids the following electro-magnetic pairs,
corresponding to the same form $F^a, a \in \tri$:

\begin{center}
\begin{tabular}{ccc}
\cline{1-2}
\multicolumn{1}{|c|}{\hspace*{1cm}} &
\multicolumn{1}{|c|}{$i$}  & $\quad\bar J$ \\
\cline{1-2}
 & & \\
\cline{1-1}
\multicolumn{1}{|c|}{\hspace*{1cm}} & & $\quad I$
\\
\cline{1-1}
\end{tabular}
\end{center}

\begin{center}

{\bf \small Figure 2.
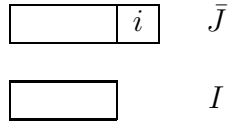
Forbidden by Restriction 2 electromagnetic pair of
p-branes}

\end{center}

Here $d_i =1$, $i =1,\dots,n$. In terms of intersections
Restriction 2 reads
\beq{2.2.3a}
{\bf (R2)} \quad d(I \cap J) \neq 0,
\eeq
for $I\in\Omega_{a,e}$ and $J\in\Omega_{a,m}$, $a \in \tri$.

\subsection{$U^s$-vectors and scalar products}

Here we remind a minisuperspace covariant form of constraints
using so called $U^s$-vectors \cite{IMC}. The constraints (\ref{1.27})
may be written in the following form
\beq{1.32}
U^s(c)= U^s_A c^A= 0, \qquad U^s(\bar c)= U^s_A \bar c^A= 0,
\eeq
where
\beq{1.33}
(U_A^s)=(d_i\delta_{iI_s},-\chi_s\lambda_{\alpha a_s}),
\eeq
$s=(a_s,v_s,I_s) \in S$, $A = (i, \alpha)$.

The quadratic constraint (\ref{1.30a}) reads
\beq{1.30b}
E=E_{TL}+ \frac12 \hat G_{AB} c^A c^B = 0,
\eeq
and
\beq{1.36}
(\hat G_{AB})=\barr{cc}
G_{ij}& 0\\
0& h_{\alpha\beta}
\earr,
\eeq
is the (truncated) target space metric with
\beq{1.37}
G_{ij}= d_i \delta_{ij} - d_i d_j,
\eeq
$i,j = 1, \ldots, n$.
Let
\beq{1.38}
(U,U')=\hat G^{AB}U_AU'_B,
\eeq
be the  scalar product,
where $U=U_Az^A$, $U' = U_A z^A$ are linear functions (1-forms)
on $\R^{n+l}$, and  $(\hat G^{AB})=(\hat G_{AB})^{-1}$.
The scalar products (\ref{1.38}) for co-vectors
$U^s$ from $(\ref{1.33})$  were calculated in
\cite{IMC}
\beq{1.39}
(U^s,U^{s'})= B_{s s'},
\eeq
$s, s' \in S$ (see (\ref{1.17})).

{\bf Intersection rules.}
>From  (\ref{1.16}), (\ref{1.17}) and (\ref{1.18})  we get
the  $p$-brane intersection rules  corresponding
to the quasi-Cartan matrix $(A_{s s'})$ \cite{IMJ}
\beq{1.40}
d(I_s \cap I_{s'})= \frac{d(I_s)d(I_{s'})}{D-2}-
\chi_s\chi_{s'}\lambda_{a_s}\cdot\lambda_{a_{s'}} + \frac12 K_{s'} A_{s s'},
\eeq
where $\lambda_{a_s}\cdot\lambda_{a_{s'}} =
\lambda_{\alpha a_s}\lambda_{\beta a_{s'}} h^{\alpha\beta}$;
$s, s' \in S$.

The contravariant components $U^{sA}= \hat G^{AB} U^s_B$ read
\cite{IMC}
\beq{1.41}
U^{si}= G^{ij}U_j^s=
\delta_{iI_s}-\frac{d(I_s)}{D-2}, \quad U^{s\alpha}= - \chi_s
\lambda_{a_s}^\alpha,
\eeq
$s \in S$.
Here  \cite{IMZ}
\beq{1.43}
G^{ij}=\frac{\delta^{ij}}{d_i}+\frac1{2-D},
\eeq
$i,j=1,\dots,n$, are the components of the matrix inverse to
$(G_{ij})$ from (\ref{1.37}). The contravariant components
(\ref{1.41}) appear as powers in relations
for the metric and scalar fields in (\ref{1.3}) and  (\ref{1.4}).

We note that the solution under consideration for the special case
of the $A_m$ Toda chain was obtained earlier  in \cite{GM2}.

\section{ Generalized fluxbrane solutions}

\subsection{The choice of parameters }

In what follows we put
\beq{2.1}
w = 1, \quad  d_1 = 1,
\eeq
i.e. the manifold $M_1$ is one-dimensional one
and
\beq{2.18}
1 \in I_s, \quad \forall s \in S,
\eeq
i.e. all branes contain $M_1$-submanifold. We note
that Restriction 2 is satisfied automatically due to
(\ref{2.18}).

Here we restrict ourselves
to  solutions with linear asymptotics at infinity
\beq{2.7}
q^s = - \beta^s u + \bar \beta^s  + o(1),
\eeq
$u \to +\infty$, where $\beta^s, \bar \beta^s$ are
constants, $s \in S$. This relation gives us an
asymptotical solution to  Toda type eqs. (\ref{1.20}) if
\beq{2.8}
\sum_{s' \in S} A_{s s'} \beta^{s'} > 0,
\eeq
for all $s \in S$. In this case the energy  (\ref{1.31})
reads
\beq{2.9}
E_{TL} = \frac{1}{4}  \sum_{s,s' \in S}
h_s A_{s s'}  \beta^s \beta^{s'}.
\eeq

We put
\beq{2.7a}
\bar{c}^A = 0,  \qquad \bar \beta^s = 0.
\eeq
$A = (i, \alpha)$, $s \in S$.
(These relations guarantee the asymptotical flatness
for $u \to +\infty$ of the $2$-dimensional section of the
metric for $M_1 = S^1$ and  $g^1 = d \phi \otimes d \phi$).

We also put
\bear{2.19}
c^A  = - \delta^{A}_{1} +
\sum_{s\in S}  h_s \beta^s U^{s A},  \\ \label{2.20}
\beta^s = 2 \sum_{s' \in S} A^{s s'},
\ear
where $s \in S$, $A = (i, \alpha)$,
and the matrix $(A^{s s'})$ is inverse to the matrix
$(A_{s s'}) = ( 2 (U^s, U^{s'})/ (U^{s'}, U^{s'}))$.
It may be verified that the constraints
(\ref{1.32}) and (\ref{1.30b}) are satisfied identically,
due to conditions (\ref{2.1}), (\ref{2.18}).

Relations (\ref{2.19}) and (\ref{2.20}) look similar to
analogous relations for black branes \cite{IMp2,IMp3,IMtop}
which appear from the horizon condition. These relations
play a key role in this paper. Their physical
sence (possibly related to regularity of the metric
at $u \to \infty$) will be clarified in a separate publication.

\subsection{The main solution}

We introduce a new radial variable
\bear{2.28}
\exp( - u) = \rho
\ear
and denote
\beq{2.28a}
H_s =
f_s e^{-  \beta^s u }= e^{- q^s -  \beta^s u },
\eeq
$s \in S$.
Then the  solutions (\ref{1.3})-(\ref{1.5}) may be written as follows
\bear{2.30}
g= \Bigl(\prod_{s \in S} H_s^{2 h_s d(I_s)/(D-2)} \Bigr)
\biggl\{ d\rho \otimes d \rho  \qquad
\\ \nn
+  \Bigl(\prod_{s \in S} H_s^{-2 h_s} \Bigr) \rho^2 g^1
+ \sum_{i = 2}^{n} \Bigl(\prod_{s\in S}
  H_s^{-2 h_s \delta_{iI_s}} \Bigr) g^i  \biggr\},
\\  \label{2.31}
\exp(\varphi^\alpha)=
\prod_{s\in S} H_s^{h_s \chi_s \lambda_{a_s}^\alpha},
\\  \label{2.32a}
F^a= \sum_{s \in S} \delta^a_{a_s} {\cal F}^{s},
\ear
where
\beq{2.32}
{\cal F}^s= - Q_s
\left( \prod_{s' \in S}  H_{s'}^{- A_{s s'}} \right)
\rho d\rho  \wedge \tau(I_s),
\eeq
$s\in S_e$,
\beq{2.33}
{\cal F}^s= Q_s \tau(\bar I_s),
\eeq
$s\in S_m$.
Here $Q_s \neq 0$, $h_s =K_s^{-1}$; parameters $K_s \neq 0$ and
the non-degenerate matrix $(A_{s s'})$ are defined by  relations
(\ref{1.16}), (\ref{1.17}), (\ref{1.18}), $s \in S$.

Functions $H_s > 0$ obey the equations
\beq{2.34}
 \frac{d}{\rho d\rho} \left(
\frac{\rho}{H_s}  \frac{d H_s}{d \rho} \right) = B_s
\prod_{s' \in S}  H_{s'}^{- A_{s s'}},
\eeq
$s \in S$, where $B_s \neq 0$ are defined in (\ref{1.21}).
These equations follow from Toda-type equations (\ref{1.20}) and
the definitions   (\ref{2.28}) and   (\ref{2.28a}).

It follows from (\ref{2.7}), (\ref{2.7a}) and  (\ref{2.28a})
that there exist finite limits
\beq{2.35}
H_s  \to 1,
\eeq
$s \in S$, for $\rho \to 0$.

For cylindrically symmetric case
\beq{2.40}
M_1 = S^1, \qquad g^1 = d \phi \otimes d \phi,
\eeq
$0 < \phi < 2 \pi$, we get a family of composite fluxbrane solutions.
They are defined up to solutions to radial equations (\ref{2.34})
with the boundary conditions  (\ref{2.35}) imposed.
In the next sections we consider several exact solutions
to eqs. (\ref{2.34}) and (\ref{2.35}).

Another possibility
\beq{2.40t}
M_1 = \R, \qquad g^1 = - d t \otimes d t,
\eeq
$- \infty < t <  + \infty$ ($t$ is time variable),
will lead us to generalized Milne-type solution (that may be
also a subject of a separate publication).

\subsection{Fluxbrane intersection rules}

The fluxbrane submanifold (worldvolume) is isomorphic to $M(\bar{I}_s)$
(see (\ref{1.13a}), (\ref{1.14})) and has the following
dimension
\beq{2.40f}
d(\bar{I}_s)  = D - 1 - d(I_s),
\eeq
$s \in S$, (see (\ref{1.16a})).
The fluxbrane intersection rules read
\beq{2.40i}
d(\bar{I}_s \cap \bar{I}_{s'})  = D - 1 -
d(I_s) - d(I_{s'}) +  d(I_s \cap I_{s'}),
\eeq
$s \neq s'$, with $p$-brane intersections
defined in (\ref{1.40}). This relation
follows from the identity
\beq{2.40s}
\bar{I}_s \cap \bar{I}_{s'} = I_0 \setminus (I_s \cup I_{s'}),
\eeq
(see (\ref{1.13a})).

We note that here we use $p$-brane notations for description
of flux $p$-branes or $Fp$-branes.
An electric (magnetic) $p$-brane corresponds to a magnetic (electric)
$F(D-3 -p)$ fluxbrane.

\section{Polynomial structure of $H_s$ for  Lie algebras}

Now we deal  with solutions to second order non-linear
differential equations  (\ref{2.34}) that may be rewritten
as follows
\beq{3.1}
 \frac{d}{dz} \left( \frac{z}{H_s} \frac{d}{dz} H_s \right) =
 \frac{1}{4} B_s
\prod_{s' \in S}  H_{s'}^{- A_{s s'}},
\eeq
where $H_s(z) > 0$, $s \in S$, $z = \rho^{2}$.
Eqs.  (\ref{2.35}) read
\beq{3.2b}
H_{s}(z = +0) = 1,
\eeq
$s \in S$.
(Here we repeat equations (\ref{0.1})-(\ref{0.2})
for convenience.)

In general one may try to seek solutions of (\ref{3.1})  in a class
of functions analytical in some disc $|z| < L$:
\beq{3.3}
H_{s}(z) = 1 + \sum_{k = 1}^{\infty} P_s^{(k)} z^k,
\eeq
where $P_s^{(k)}$ are constants, $s \in S$.
Substitution
of (\ref{3.3})  into (\ref{3.1}) gives us an infinite
chain of relations on parameters $P_s^{(k)}$  and
$B_s$.  The first relation in this chain
\beq{3.5a}
P_s  \equiv  P_s^{(1)} = \frac{1}{4} B_s,
\eeq
$s \in S$, corresponds to the $z^0$-term in the decomposition
of (\ref{3.1}). For analytical function $H_s(z)$
(\ref{3.3}) ($z = \rho^2$) the metric (\ref{2.30}) is regular at
$\rho = 0$.

\subsection{Solutions for Lie algebra
$A_1 \oplus  \ldots  \oplus  A_1$}

\subsubsection{Orthogonal case}

Meanwhile there exist solutions to eqs. (\ref{3.1})-(\ref{3.2b})
of polynomial type. The simplest example occurs in orthogonal
case, when
\beq{3.4}
(U^s,U^{s'})= B_{s s'} = 0,
\eeq
for  $s \neq s'$, $s, s' \in S$. In this case
$(A_{s s'}) = {\rm diag}(2,\ldots,2)$ is a Cartan matrix
for semisimple Lie algebra
$A_1 \oplus  \ldots  \oplus  A_1$
and
\beq{3.5}
H_{s}(z) = 1 + P_s z,
\eeq
with $P_s \neq 0$ satisfying (\ref{3.5a}). (For $A_1$-case
see  \cite{CGSaf}).

\subsubsection{Block-orthogonal case}

The  solution (\ref{3.5})
may be  generalized to so-called
"block-orthogonal" (BO) case:
\beq{3.6}
S=S_1 \sqcup \dots \sqcup S_k,
\eeq
$S_i \ne \emptyset$, i.e. the set $S$ is a union of $k$ non-intersecting
(non-empty) subsets $S_1,\dots,S_k$,
and
\beq{3.7}
(U^s,U^{s'})=0
\eeq
for all $s\in S_i$, $s'\in S_j$, $i\ne j$; $i,j=1,\dots,k$.
(For "block-orthogonal" black branes see \cite{Br,CIM}.)
In this case (\ref{3.5}) is modified as follows
\beq{3.8}
H_{s}(z) = (1 + {\bar P}_s z)^{\beta^s},
\eeq
where $\beta^s$ are defined in  (\ref{2.20})
and parameters $ {\bar P}_s$   are coinciding inside
blocks, i.e. ${\bar P}_s = {\bar P}_{s'}$
for $s, s' \in S_i$, $i =1,\dots,k$,
and satisfy the following relations
\beq{3.5b}
    {\bar P}_s =  B_s/(4 \beta^s),
\eeq
for $\beta^s \neq 0$, $s \in S$. In this case $H_s$
are analytical in $|z| < L$, where $L =
{\rm min} (| {\bar P}_ s|^{-1}, s \in S$).

Let $(A_{s s'})$ be  a Cartan matrix  for a  finite-dimensional
semisimple Lie  algebra $\cal G$. In this case all powers in
(\ref{2.20})  are  natural numbers
\beq{3.11}
\beta^s = 2 \sum_{s' \in S} A^{s s'} = n_s \in \N,
\eeq
and  hence, all functions $H_s$ are polynomials, $s \in S$.
Integers $n_s$ coincide with the components  of the twice
dual Weyl vector in the basis of simple roots \cite{FS}.

{\bf Conjecture.} {\em Let $(A_{s s'})$ be  a Cartan matrix
for a  semisimple finite-dimensional Lie algebra $\cal G$.
Then  the solution to eqs. (\ref{3.1}), (\ref{3.2b})
(if exists) is a polynomial
\beq{3.12}
H_{s}(z) = 1 + \sum_{k = 1}^{n_s} P_s^{(k)} z^k,
\eeq
where $P_s^{(k)}$ are constants,
$k = 1,\ldots, n_s$, integers $n_s$ are
defined in (\ref{3.11}) and $P_s^{(n_s)} \neq 0$,  $s \in S$}.

This conjecture may be verified for Lie algebras
$A_m$, $C_{m+1}$ repeating all arguments from
\cite{IMp2,IMp3}  with the replacement of $F(z) = 1 - 2 \mu z $
by $F(z) = z$.

\subsection{Solutions for Lie algebra $A_2$}

Let us consider the Lie algebra $A_2 = sl(3)$ with the Cartan matrix
\beq{4.5}
    \left(A_{ss'}\right)=
  \left( \begin{array}{*{6}{c}}
     2 & -1\\
     -1& 2\\
\end{array}
\right)\quad
\eeq
According
to the {\bf Conjecture} we  seek the solutions
to eqs. (\ref{3.1})-(\ref{3.2b}) in the following
form ($n_1 = n_2 =2$):
\beq{4.6}
H_{s} = 1 + P_s z + P_s^{(2)} z^{2},
\eeq
where $P_s$ and $P_s^{(2)} \neq 0$ are constants,
$s = 1,2$.

The substitution of  (\ref{4.6}) into equations  (\ref{3.1})
and decomposition in powers of $z$ lead us to relations
(\ref{3.5a}) and
\bear{4.7}
   P_s^{(2)}  =  \frac14 P_1 P_2.
\ear

Thus, in the $A_2$-case the solution is described by relations
(\ref{2.30})-(\ref{2.33}) with $S = \{s_1,s_2\}$,
$p$-brane intersection rules (\ref{1.40}), or, equivalently,
\bear{1.40a}
d(I_{s_1} \cap I_{s_2})= \frac{d(I_{s_1})d(I_{s_2})}{D-2}-
\chi_{s_1} \chi_{s_2} \lambda_{a_{s_1}}\cdot\lambda_{a_{s_2}}
- \frac12 K,
\\ \label{1.40b}
d(I_{s_i}) - \frac{(d(I_{s_i}))^2}{D-2}+
\lambda_{a_{s_i}}\cdot\lambda_{a_{s_i}} = K,
\ear
where
$K = K_{s_i} \neq 0$,  and functions $H_{s_i} = H_i$
are defined by relations
(\ref{4.6}), (\ref{3.5a}), (\ref{4.7}) with $z = \rho^2$,
$i =1,2$.

\section{Examples of fluxbrane solutions}

Here we present certain examples of fluxbrane solutions with
$M_1 = S^1$ and $g^1 = d \phi \otimes d \phi$.
In all examples below the total metric $g$ has the signature
$(-,+,..., +)$ and all ($p$-brane) signature parameters
are positive: $\eps_s = \eps(I_s)= +1$ (here all $\theta_a =1$).
In what follows $0 < \rho < +\infty$.  In
all examples the metrics are regular at $\rho = 0$.

\subsection{Solutions for algebra $A_1$}

We start with single fluxbrane solutions ($S = \{ s \}$).

\subsubsection{Melvin  solution ($F1$ fluxbrane) }

Let  $D= 4$, $n =3$, $M_2 = \R$ with
$g^2 = - dt \otimes dt$ and $M_3 = \R$ with  $g^3 = d \eta \otimes d\eta$,
and $I_s  = \{ 1 \}$.
The solution reads \cite{Melv}
\bear{4.1a}
g=  H^2 \biggl\{ d\rho \otimes d \rho +
H^{-4} \rho^2 d \phi \otimes d \phi
- dt \otimes dt + d \eta \otimes d \eta \biggr\}.
\\  \label{4.1b}
F = - Q H^{- 2} \rho d\rho \wedge d \phi,
\ear
where $H = 1 + \frac{1}{8} Q^2 \rho^2 $.
Here $-Q$ is proportional to magnetic field in the core.

\subsubsection{$F6$ fluxbrane (corresponding to $M2$-brane)}

Consider $D =11$ supergravity with the metric $g$ and $4$-form $F$
in the bosonic sector \cite{CJS}.
Let $n =3$, $M_3$ be $7$-dimensional (Ricci-flat) manifold with the
metric $g^3 = g^3_{\mu \nu} dx^{\mu} \otimes dx^{\nu}$
of signature $(-,+, \ldots, +)$ and $M_2$ be $2$-dimensional (flat)
manifold  of signature $(+,+)$ with the
metric $g^2 = g^2_{m n} dy^m \otimes dy^n$ and $I_s = \{1, 2\}$.
The solution reads
\bear{4.2}
g=  H^{1/3} \biggl\{ d\rho \otimes d \rho
+  H^{-1} ( \rho^2 d \phi \otimes d \phi + g^2) + g^3   \biggr\},
\\  \label{4.2a}
F= - Q H^{- 2} \rho d\rho  \wedge d \phi \wedge \tau_2,
\ear
where
$H = 1 + \frac{1}{2} Q^2 \rho^2 $.
For flat $g^3$ this solution was obtained earlier in
\cite{CGS}.

\subsubsection{$F3$ fluxbrane (corresponding to $M5$-brane)}

Now we  consider the solution dual to $F6$.
Let $n =3$, $M_3$ be $4$-dimensional (Ricci-flat) manifold with the
metric $g^3 = g^3_{\mu \nu} dx^{\mu} \otimes dx^{\nu}$
of signature $(-,+, +, +)$,
and $M_2$ be $5$-dimensional (Ricci-flat) manifold
of signature $(+, \ldots, +)$ with the
metric $g^2 = g^2_{m n} dy^m \otimes dy^n$ and $I_s = \{1, 2\}$.
The solution reads
\bear{4.3}
g=  H^{2/3} \biggl\{ d\rho \otimes d \rho
+  H^{-1} ( \rho^2 d \phi \otimes d \phi + g^2) + g^3   \biggr\},
\\  \label{4.3a}
F= Q \tau_3,
\ear
where $H = 1 + \frac{1}{2} Q^2 \rho^2 $.
For flat $g^2$ and $g^3$ see \cite{CGS}.

\subsubsection{$F7,F6,F5$ and $F1, F2, F3$ fluxbranes
in $IIA$ supergravity}

The bosonic part of action
for $D=10$ $IIA$ supergravity reads
\beq{4.3.s1}
S=\int d^{10}z\sqrt{|g|}\biggl\{R[g]-(\p\varphi)^2-\sum_{a=2}^4
e^{2\lambda_a\varphi}(F^a)^2\biggr\}-\frac12\int F^4 \wedge F^4\wedge A^2,
\eeq
where $F^a=dA^{a-1}+\delta^a_4 A^1 \wedge F^3$ is an $a$-form,
$a=2,3,4$, and
\beq{4.3.s2}
\lambda_2=3\lambda_4, \qquad \lambda_3=-2\lambda_4, \qquad
\lambda_4^2=  1/8.
\eeq
The dimensions of $p$-brane worldvolumes are
\beq{4.3.s3}
d = d(a, \chi) =  \left\{\begin{array}{ll}
1,2,3&\mbox{ in electric case}, \quad \chi = +1, \\
7,6,5&\mbox{ in magnetic case}, \quad \chi = -1,
\end{array}\right.
\eeq
for $a=2,3,4$, respectively.

The fluxbrane solutions read
\bear{4.3.s4}
g=  H^{d/8} \biggl\{ d\rho \otimes d \rho
+   H^{-1}  \rho^2 d \phi \otimes d \phi
+   H^{-1}  g^2  +  g^3  \biggr\},
\\  \label{4.3.s5}
\exp(2 \varphi)=
H^{ \chi \lambda_a},
\ear
and
\beq{4.3.s7}
F^a= - Q  H^{- 2} \rho d\rho  \wedge \tau_2,
\quad {\rm for} \quad \chi = +1,
\eeq
or
\beq{4.3.s8}
F^a= Q \tau_3,  \quad {\rm for} \quad \chi = - 1,
\eeq
with $H = 1 + \frac{1}{2} Q^2 \rho^2 $, $a=2,3,4$.

For $\chi = +1$  we get a $F(9-a)$ fluxbrane with
$d_2 = a -2$, $d_3 = 10-a$,  and
for $\chi = -1$  we obtain a (dual) $F(a-1)$ fluxbrane  with
$d_2 = 8-a$, $d_3 = a$,  ($a=2,3,4$). (Here the presence of
Chern-Simons terms does not modify the solutions from Sect. 3.2.)

\subsection{Solution for  algebra $A_1 \oplus A_1$ }

{\bf $F6 \cap F3$ fluxbranes.}
We put $n =5$, $d_2 =1$, $g^2 = dy^2 \otimes dy^2$,
$d_3 =1$, $g^3 = dy^3 \otimes dy^3$,
$d_4 =4$ and $g^4$ has the Euclidean signature,
$d_5 =3$ and $g^5$ has the signature $(-,+, +)$,
$I_{s_e} = \{1,2,3 \} $ and $I_{s_m} = \{1,2,4 \}$.
The solution has the following form
\bear{4.4}
g=  H_e^{1/3} H_m^{2/3} \biggl\{ d\rho \otimes d \rho
+  H_e^{-1} H_m^{-1} ( \rho^2 d \phi \otimes d \phi + g^2) +
H_e^{-1} g^3 + H_m^{-1} g^4 + g^5  \biggr\},
\\  \label{4.4a}
F= - Q_e H_e^{- 2} \rho d\rho  \wedge d \phi \wedge dy_2 \wedge dy_3 +
Q_m dy_3 \wedge \tau_5,
\ear
where $H_s = 1 + \frac{1}{2} Q_s^2 \rho^2 $, $s = e,m$.
For flat $g^i$ see  \cite{CGS}.

\subsection{Solutions for algebra $A_2$}

\subsubsection{$F6 \cap F3$ fluxbranes with $A_2$-intersection}

Now we consider a new  $F6 \cap F3$ fluxbrane configuration
with (a non-standard) $A_2$ intersection rules
defined on the manifold
\beq{4.8}
M =    (0, +\infty )
 \times M_1 \times M_{2} \times M_{3} \times M_4,
\eeq
where $d_2 =  2$, $d_3 =  5$, $d_4 =2$.
The solution is following
\bear{4.9}
g=  H_e^{1/3} H_m^{2/3} \biggl\{ d\rho \otimes d\rho +
%\\ \nn
   H_e^{-1} H_m^{-1} \rho^2 d \phi \otimes d \phi
+ H_e^{-1} g^2 + H_m^{-1} g^3  + g^4 \biggr\}, \mm
\label{4.10}
F =  - Q_e H_e^{-2} H_m  \rho d\rho\wedge d\phi \wedge \tau_2+
Q_m \tau_2 \wedge \tau_4,
\ear
where metrics $g^2$ and  $g^3$ are
(Ricci-flat) metrics of Euclidean signature,
$g^4$ is the (flat) metric of the signature $(-,+)$
and
\beq{4.11}
H_{s} = 1 + P_s \rho^2+  \frac{1}{4} P_1 P_2 \rho^4,
\eeq
where  $P_s = \frac{1}{2} Q_s^2$, $s =e,m$.

\subsubsection{Dyonic flux tube in Kaluza-Klein model}

Let us  consider $4$-dimensional model
\beq{4.15}
S= \int_{M} d^4z \sqrt{|g|}\biggl\{R[g]- g^{\mu \nu}
\p_\mu \varphi \p_\nu \varphi
-\frac{1}{2!} \exp[2\lambda \varphi]F^2\biggr\}
\eeq
with scalar field  $\varphi$, two-form $F = d A$ and
$\lambda = - \sqrt{3/2}$.
This model originates after Kaluza-Klein reduction
of $5$-dimensional gravity. The 5-dimensional metric
in this case reads
\beq{4.16a}
g^{(5)} = \Phi g_{\mu \nu} dx^{\mu} \otimes dx^{\nu}
            + \Phi^{-2} (dy + {\cal A}) \otimes (dy + {\cal A}),
\eeq
where
\beq{4.16b}
{\cal A} = \sqrt{2} A =  \sqrt{2} A_{\mu} dx^{\mu},
\qquad \Phi = \exp(2 \varphi/\sqrt{6}).
\eeq

We consider  "dyonic" flux-tube solution
defined on the manifold
\beq{4.17}
M =    (0, +\infty )  \times (M_1 =S^{1})  \times M_2,
\eeq
where $M_2 = \R^2$ and $g^2 = - dt \otimes dt + d \eta \otimes d \eta$.
This solution reads
\bear{4.18}
g= \left( H_e H_m \right)^{1/2}
   \biggl\{ d \rho \otimes d\rho
 - H_e^{-1} H_m^{-1}  \rho^2 d \phi \otimes d \phi
- dt \otimes dt + d \eta \otimes d \eta \biggl\},
\\ \label{4.19}
\exp \varphi = H_e^{\lambda/2} H_m^{- \lambda/2},
\\ \label{4.20}
F = dA = -  Q_e  H_e^{-2} H_m \rho d\rho \wedge d\phi +
Q_m dt \wedge d \eta,
\ear
where  functions $H_{s}$  are defined by relations (\ref{4.11}),

For 5-metric we obtain from (\ref{4.16a})-(\ref{4.19})
\bear{4.16c}
g^{(5)} = H_m \{ d\rho \otimes d\rho
+  H_e^{-1} H_m^{-1} \rho^2 d\phi \otimes d \phi
- dt \otimes dt + d \eta \otimes d \eta \}
\\ \nn
 + H_e H_m^{-1}  (dy + {\cal A}) \otimes (dy + {\cal A}),
\ear
$d {\cal A} = \sqrt{2} F$.
For $Q_m \to 0$ we get the  solution from \cite{GM}
and for $Q_e \to 0$ we are led to its dual version (see \cite{CGS}).

\section{Conclusions}

Thus, here we obtained a family of fluxbrane solutions  with general
intersection rules. (See relations (\ref{1.17a}), (\ref{1.18b}) and
Restriction 1. Restriction 2 is satisfied, since all
$p$-branes have a common $M_1$.) These  solutions are given by relations
(\ref{2.30})-(\ref{2.35}).
The metrics of solutions  contain  $n$ Ricci-flat metrics.
The solutions are defined up to  a set of
"moduli" functions $H_s$  obeying a set of  equations
(\ref{2.34}) with  the boundary conditions (\ref{2.35})  imposed.
These solutions are new and generalize a lot of
special fluxbrane  solutions considered earlier in the literature.

Here we suggested a conjecture on polynomial structure of  $H_s$
for intersections related to semisimple Lie algebras.
This conjecture is valid for Lie algebras $A_m$ and $C_{m+1}$,
$m \geq 1$, at may be verified  with a little modification of the proof
for the black brane case \cite{IMp2,IMp3}.

We obtained explicit formulas for $A_1 \oplus \ldots \oplus A_1$
(orthogonal), block-ortogonal and $A_2$ solutions.
These formulas are illustrated by certain
examples of solutions in  $D = 10, 11$
supergravities (e.g. with $A_2$ intersection
rules) and  Kaluza-Klein dyonic $A_2$ flux tube.
Explicit relations  for $H_s$ corresponding
to other examples of Lie algebras (e.g.  $A_3$, $B_2$ etc) will be
a subject of further publications.
(Another topic of interest is related to
supersymmetric  composite fluxbranes on product of
Ricci-flat manifolds. For $M$-branes see \cite{Iv2} and references
therein).

\begin{center}
{\bf Acknowledgments}
\end{center}

This work was supported in part by the Russian Ministry for
Science and Technology, Russian Foundation for Basic Research,
DFG foundation and project SEE.

\end{document}